\documentclass[11pt,review]{elsarticle}
\biboptions{numbers,sort&compress} 
\journal{arXiv}
\usepackage{hyperref}
\usepackage{lineno}

\usepackage[utf8]{inputenc}
\usepackage{graphicx}
\usepackage{booktabs} 
\usepackage{xcolor}
\usepackage{amsmath, amssymb, amsthm} 
\allowdisplaybreaks
\newcommand{\TABREF}[1]{Table~\ref{#1}}

\makeatletter
\let\@afterindenttrue\@afterindentfalse
\makeatother

\begin{document}

\begin{frontmatter}

\title{Shadow Prices and Optimal Cost in Economic Applications}

\author[addr-iiasa-em]{Nikolay Khabarov\corref{tag-correspondingauthor}}
\cortext[tag-correspondingauthor]{Corresponding author}
\ead{khabarov@iiasa.ac.at} 

\author[addr-iiasa-ibf,addr-msu]{Alexey Smirnov}

\author[addr-iiasa-em,addr-oxford]{Michael Obersteiner}

\address[addr-iiasa-em]{Exploratory Modeling of Human-Natural Systems Research Group, 
Advancing Systems Analysis Program,
International Institute for Applied Systems Analysis (IIASA), 
Schlossplatz 1, Laxenburg, A-2361, Austria}

\address[addr-iiasa-ibf]{Integrated Biosphere Futures Research Group,  Biodiversity and Natural Resources Program,
International Institute for Applied Systems Analysis (IIASA), 
Schlossplatz 1, Laxenburg, A-2361, Austria}

\address[addr-msu]{Faculty of Computational Mathematics and Cybernetics,
Lomonosov Moscow State University, Moscow, 119991, Russia}

\address[addr-oxford]{Environmental Change Institute, Oxford University Centre for the Environment, South Parks Road, Oxford, OX1 3QY, UK}

\begin{abstract}

Shadow prices are well understood and are widely used in economic applications. 
However, there are limits to where shadow prices can be applied assuming their natural interpretation and the fact that they reflect the first order optimality conditions (FOC). 
In this paper, we present a simple ad-hoc example demonstrating that marginal cost associated with exercising an optimal control may exceed the respective cost estimated from a ratio of shadow prices. 
Moreover, such cost estimation through shadow prices is arbitrary and depends on a particular (mathematically equivalent) formulation of the optimization problem.
These facts render a ratio of shadow prices irrelevant to estimation of optimal marginal cost. 
The provided illustrative optimization problem links to a similar approach of calculating social cost of carbon (SCC) in the widely used dynamic integrated model of climate and the economy (DICE).
\end{abstract}


\end{frontmatter}

\section{Introduction}

The concept of shadow prices is widely known and has extensive applications in economics. There is a large body of literature devoted to exploration of properties and limitations of shadow prices in various settings e.g. in competitive general equilibrium models~\cite{refSmith1987}, valuation of nonmarket goods~\cite{refStarrett2000}, industrial pollutants~\cite{refDanWu2021}, environmental efficiency~\cite{refElvira2022}, carbon pricing and abatement costs~\cite{refZhaohua2022}, land use regulation~\cite{refJunfu2022}, land allocation~\cite{refAslihan2011}, cost of soil erosion~\cite{refXiaojie2022}, patent and knowledge stock valuation~\cite{refMichiyuki2018}, water supply~\cite{refAlexandros2020}, power system reliability assessment~\cite{refKai2022}, portfolio optimization~\cite{refFrancesca2019}, social cost of carbon (SCC) estimation~\cite{refNordhaus2019},  ~\cite{refDICEWebPage}. 

In economic applications formulated as a constrained optimization problem, a shadow price is the change in the optimal value of the objective function (e.g. utility) per unit of a constraint that is relaxed by an infinitesimal amount~p.452 in~\cite{refSimonBlume1994}. In the context of continuous time optimal control problems, shadow price depends on time and is referred to as a  “costate variable”~\cite{refFeichtinger2008}.

In this paper, we focus on a non-linear constrained optimization problem similar to DICE~\cite{refNordhaus2019},  ~\cite{refDICEWebPage} and extend the analysis~\cite{refSCCFrontiers} to further explore the method of estimating optimal cost based on shadow prices. 
Such method involving a ratio of two shadow prices (or marginal values) is employed in DICE for estimation of the social cost of carbon.
A similar approach of deriving a crop price as a ratio of two shadow prices is suggested in~\cite{refAslihan2011}. The possibility of such derived price to deviate from an (exogenous) market price is admitted in that paper, whereas explanation of this deviation is suggested to be in the area of non-market values.

The main driving question for the following analysis is whether the cost \emph{estimate} of an optimal action obtained via shadow prices would be equal to the \emph{actual} cost corresponding to an optimal solution.
To answer that question, we (1) construct a simple utility maximization problem, (2) solve that problem numerically, (3) check first-order optimality conditions to ensure high accuracy of the obtained solution, (4) compare the cost of optimal control versus its cost estimate obtained via shadow prices, and (5) demonstrate that the latter is an underestimate of the former.
We further illustrate that the cost estimated through shadow prices is an arbitrary number, because it depends on a particular (mathematically fully equivalent) formulation of the optimization problem and can be lower, equal, or higher than the optimal cost.
We finally conclude that despite shadow prices are derived from the first-order optimality conditions, the cost estimate obtained as a ratio of two shadow prices is, generally speaking, a not optimal arbitrary number and, as such, is of no use in applications.

\section{Optimization problem formulation}

We consider a hypothetical fruit storage facility operating in two time periods $i = \{1, 2\}$ and in these periods carrying $p_i$ ton of product (e.g. apples).
The storage facility manager has to protect apples from fruit flies and in doing so they decide on the pesticide application rate $\mu \in [0, 1]$.
Apples, not fully treated with pesticides support flies' reproduction and so virtually "emit" $e_i$ flies, described by the flies emission rate $\sigma$ per ton of untreated apples.
Flies in quantity $e_i$ produce damage to the stored product, which is described by the share $d(e_i)$ of stored apples that are damaged by flies.
Pesticides treatment cost is described by the cost function $g(\mu)$.
The share of product saved from consumption (selling) in period 1 is described by the saving ratio $s$, which is a control variable along with $\mu$.
There is no pesticides treatment in the period 2 and all product left from period 1 (less the share damaged by flies) is consumed (sold) in period 2.
Flies are completely removed after the end of the 1st period and before beginning of the 2nd period, so that $e_1$ does not have impact on the 2nd period.
The objective of the fruit storage facility manager is to maximize utility, which depends on consumption $c_i$ in each of the two time periods.
Mathematical formulation of the problem is presented below:
\begin{eqnarray}
	\mbox{maximize}_{\mu, s} && \sqrt{c_1} + \sqrt{c_2}, \label{bkm:eq1}  
	\\ c_1 &=& [1 - d(e_1)] p_1 (1 - s) - p_1 g(\mu), 
	\\ e_1 &=& \sigma p_1 (1 - \mu), 
	\\ p_1 &=& \mbox{const}, 
	\\ p_2 &=& [1 - d(e_1)] p_1 s, 
	\\ c_2 &=& [1 - d(e_2)] p_2, 
	\\ e_2 &=& \sigma p_2,
	\\ && 0 \le \mu \le 1, \quad 0 \le s   \le 1,
	\\ && \mu, s: \quad c_1 \ge 0, \: c_2 \ge 0. \label{bkm:eq_last}
\end{eqnarray}

\section{Optimal cost of control and its estimate from shadow prices}

Let's consider the problem (\ref{bkm:eq1}) -- (\ref{bkm:eq_last}) for a simple set of functions and constants:
\begin{eqnarray}
	d(e_i) &=& e_i,
	\\ p_1 &=& 1,
	\\ g(\mu) &=& \mu^2,
	\\ \sigma &=& 1.
\end{eqnarray}
So the problem (\ref{bkm:eq1}) -- (\ref{bkm:eq_last}) can be formulated as
\begin{eqnarray}
	\mbox{maximize}_{\mu, s} && \sqrt{c_1} + \sqrt{c_2}, \label{bkm:eq_compact1}  
	\\ c_1 &=& (1 - e_1) (1 - s) - \mu^2, \label{bkm:eq_compact2}
	\\ e_1 &=& 1 - \mu, \label{bkm:eq_compact3} 
	\\ c_2 &=& (1 - e_2) \mu s, \label{bkm:eq_compact4}
	\\ e_2 &=& \mu s, \label{bkm:eq_compact5}
	\\ && 0 \le \mu \le 1, \quad 0 \le s   \le 1, \label{bkm:eq_dom_1}
	\\ && \mu (1 - s) - \mu^2 \ge 0. \label{bkm:eq_compact_last} \label{bkm:eq_dom_2}
\end{eqnarray}


The problem (\ref{bkm:eq_compact1}) -- (\ref{bkm:eq_compact_last}) can be rewritten as 
\begin{eqnarray}
	&& \mbox{maximize}_{\mu, s} \quad f(\mu, s) \label{bkm:eq_one)line_prob_start}
	\\ && f(\mu, s) = \sqrt{\mu (1 - s) - \mu^2} 
	+ \sqrt{(1 - \mu s) \mu s}, \label{bkm:eq_one_line_prob}  
	\\ && 0 \le \mu \le 1, \quad 0 \le s   \le 1, \label{bkm:eq_one_lene_prob_dom_1}
	\\ && 1 - s \ge \mu. \label{bkm:eq_one_lene_prob_dom_2}
\end{eqnarray}

\begin{figure}[h]
  \centering
  \includegraphics[width=10 cm]{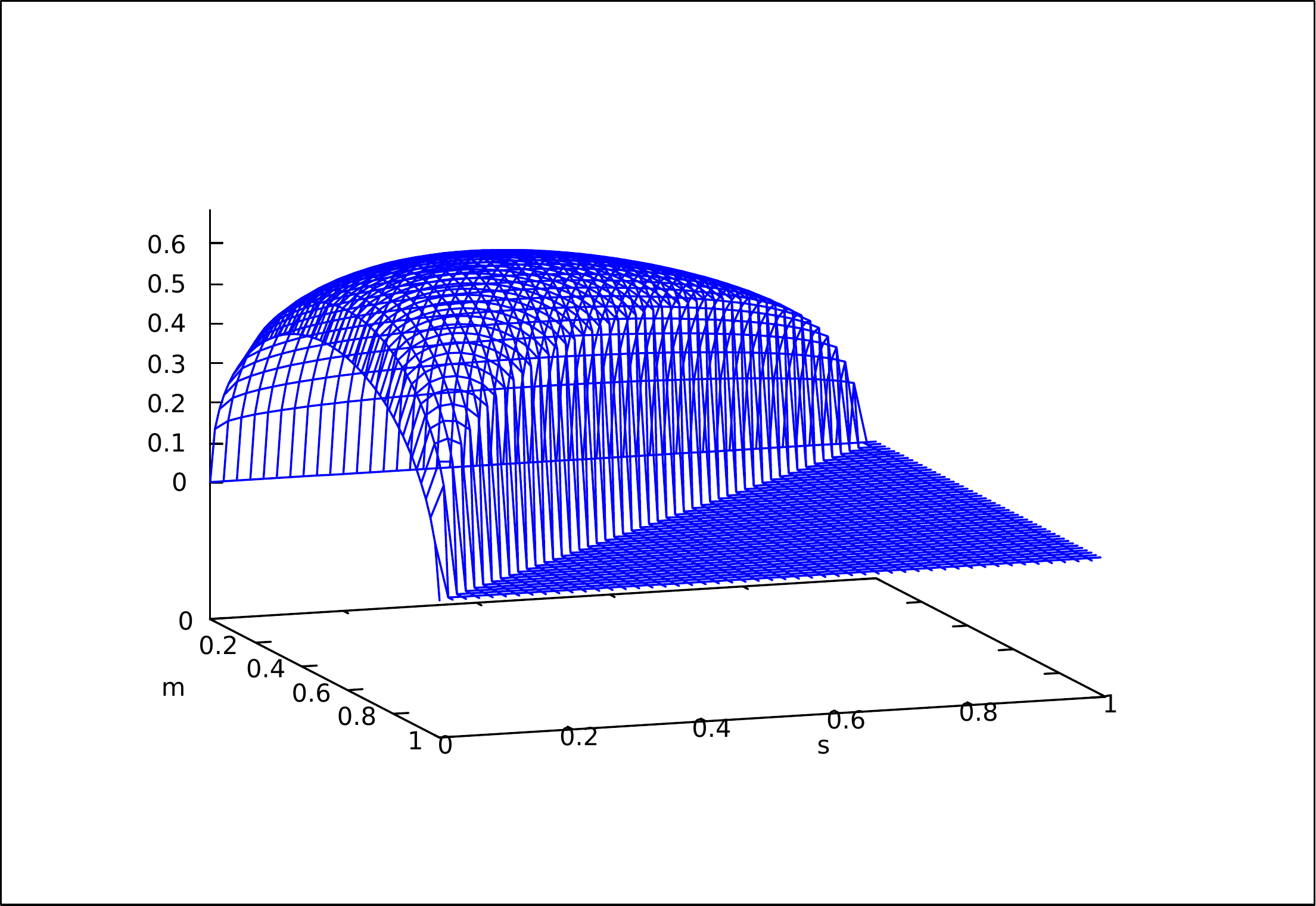}
	\caption{Function $f(\mu,s)$ in the maximization problem (\ref{bkm:eq_one)line_prob_start}) - (\ref{bkm:eq_one_lene_prob_dom_2}), here we denoted $\mu \equiv m$.}
  \label{bkm:fig1}
\end{figure}

The function $f(\mu,s)$ is strictly concave in the specified domain (\ref{bkm:eq_one_lene_prob_dom_1}), (\ref{bkm:eq_one_lene_prob_dom_2}) as illustrated by the plot of this function in Figure~\ref{bkm:fig1}. Therefore, there is a unique solution to the problem (\ref{bkm:eq_compact1}) -- (\ref{bkm:eq_compact_last}) which is an internal point of the domain (\ref{bkm:eq_one_lene_prob_dom_1}), (\ref{bkm:eq_one_lene_prob_dom_2}) or, equally, of the domain (\ref{bkm:eq_dom_1}), (\ref{bkm:eq_dom_2}). 

Let's consider an internal maximum point of the problem (\ref{bkm:eq_compact1}) -- (\ref{bkm:eq_compact_last}) i.e. when the constraints of the type of inequality (i.e. (\ref{bkm:eq_dom_1}) and (\ref{bkm:eq_dom_2})) are strictly satisfied. In this case Lagrange multipliers corresponding to inequality constraints are all zeros, and the first order optimality conditions (FOC) can be expressed through the Lagrange function $L$ employing only four multipliers $\lambda_1$, $\lambda_2$, $\xi_1$, and $\xi_2$ as follows:
\newpage 
\begin{eqnarray}
	L = - \sqrt{c_1} - \sqrt{c_2} &-& \lambda_1[ (1 - e_1) (1 - s) - \mu^2 - c_1 ] \nonumber
	\\ &-& \xi_1[ 1 - \mu - e_1 ] \nonumber
	\\ &-& \lambda_2 [ (1 - e_2) \mu s - c_2 ] \nonumber
	\\ &-& \xi_2 [ \mu s - e_2 ], \label{bkm:eq_lagr_func}
\end{eqnarray}
\begin{eqnarray}
	&& \frac{\partial L}{\partial \mu} = 2 \lambda_1 \mu + \xi_1 - \lambda_2 (1 - e_2) s - \xi_2 s = 0, \label{bkm:eq_foc1}
	\\ && \frac{\partial L}{\partial s} = \lambda_1 (1 - e_1) - \lambda_2 (1 - e_2) \mu - \xi_2 \mu = 0,
	\\ && \frac{\partial L}{\partial c_1} = - \frac{1}{2 \sqrt{c_1}} + \lambda_1 =0, \quad \frac{\partial L}{\partial c_2} = - \frac{1}{2 \sqrt{c_2}} + \lambda_2 =0,
	\\ && \frac{\partial L}{\partial e_1} = \lambda_1 (1 - s) + \xi_1 = 0, \quad \frac{\partial L}{\partial e_2} = \lambda_2 \mu s + \xi_2 = 0. \label{bkm:eq_foc_last}
\end{eqnarray}
The problem (\ref{bkm:eq_compact1}) -- (\ref{bkm:eq_compact_last}) was solved numerically using GAMS\footnote{https://www.gams.com} and all equations and inequalities (\ref{bkm:eq_compact2}) -- (\ref{bkm:eq_compact_last}) have been checked manually along with the FOC (\ref{bkm:eq_foc1}) -- (\ref{bkm:eq_foc_last}) and found to be all satisfied with high precision in the order of $10^{-8}$. Inequality constraints (\ref{bkm:eq_one_lene_prob_dom_1}), (\ref{bkm:eq_one_lene_prob_dom_2}) and, equally, (\ref{bkm:eq_dom_1}), (\ref{bkm:eq_dom_2}) are strictly satisfied: $\mu=0.5$, $s=0.206$. This indicates that the solution found numerically has high precision.

According to the meaning of the Lagrange multipliers~\cite{refSimonBlume1994}, $\lambda_1$ and $\xi_1$ are increments in the optimal value of the objective function (\ref{bkm:eq_compact1}) when one unit is added to the right hand side of the equations (\ref{bkm:eq_compact2}) and (\ref{bkm:eq_compact3}) respectively.
Let's pose the question on how many units of consumption $x$ (expressed in apples or equally in dollars) would need to be added to the consumption equation in the first period (\ref{bkm:eq_compact2}) to compensate one unit of flies added to the equation (\ref{bkm:eq_compact3}) so that the resulting value of the objective function (\ref{bkm:eq_compact1}) would not change. 
The answer is provided by the following equation relying on the meaning of $\lambda_1$ and $\xi_1$ (where increment of the objective function corresponding to $\lambda_1$ is scaled to $x$ units)
\begin{eqnarray}
	x \lambda_1 + \xi_1 = 0. \label{bkm:eq_margInc}
\end{eqnarray}
This equation holds for "sufficiently small" $\lambda_1$ and $\xi_1$ with increasingly high precision\footnote{In mathematical terms, that means that the absolute deviation of the left-hand side of the equation from zero is constrained as $|x\lambda_1 + \xi_1| \le \gamma (\lambda_1^2 + \xi_1^2)$ for some constant $\gamma > 0$.} and states that the added quantities of one unit of "flies" and $x$ apples (dollars) lead together to zero increment in the optimal value of the objective function (utility). 
So, $x$ dollars compensate one unit of "flies", i.e. $x$ can be seen as the price compensating one ton of more flies and is expressed as
\begin{equation}
	x \: = \: - \frac{\xi_1}{\lambda_1} \: = \: 0.794 \quad \mbox{(\$/ton)}. \label{bkm:eq_scc}
\end{equation}
Despite this kind of "cost estimate" $x$ (cost incurred by one ton of flies) is calculated using optimal values $\lambda_1$ and $\xi_1$, it is not the optimal cost a decision maker would pay at the optimum for reducing one ton of flies, which is expressed as
\begin{equation}
	- \left. \left. \frac{d g(\mu)}{d \mu} \right/ 
	\frac{d e_1(\mu)}{d \mu} \right|_{\mu=0.5} \: = \: 1.000 \quad \mbox{(\$/ton)}. \label{bkm:eq_smac}
\end{equation}
The minus sign in front of the equation serves the purpose of having the cost expressed as a positive number since $d e_1(\mu)/d \mu < 0$. 
From (\ref{bkm:eq_scc}) and (\ref{bkm:eq_smac}) it can be concluded that equation (\ref{bkm:eq_margInc}) along with the shadow prices $\lambda_1$ and $\xi_1$ do not define the optimal pesticide treatment cost (expressed in dollars per ton of flies reduction).

\section{Arbitrariness of cost estimate from shadow prices}

It turns out that the value of the cost estimate from shadow prices depends on which of the fully equivalent definitions of the problem (\ref{bkm:eq_compact1}) -- (\ref{bkm:eq_compact_last}) is used for such estimate. 
\TABREF{bkm:tbl1} presents the respective values obtained from equivalent optimization problems where one of the equations (\ref{bkm:eq_compact4}) or (\ref{bkm:eq_compact5}) (or both) were substituted with their respective equivalents: 
\begin{align}
	& c_2 = (1 - e_2) (1 - e_1) s, \tag{17.a}
	\\ & e_2 = (1 - e_1) s.  \tag{18.a}
\end{align}
As evidenced by these results, depending on a particular way of specifying the same problem, there are cases when the cost estimate is less than, greater than, or equal to the optimal cost (\ref{bkm:eq_smac}). This fact is easily explained by the meaning of Lagrange multipliers (marginal values to equations), which are linked to a perturbed problem that substantially depends on a particular equivalent formulation of the original problem.

\begin{table}[h]
	\caption{Optimal cost estimates obtained as a ratio of shadow prices from four equivalent formulations of the original problem (\ref{bkm:eq_compact1}) -- (\ref{bkm:eq_compact_last}) where alternative equation formulations (17.a) and (18.a) were optionally replacing the original formulations (\ref{bkm:eq_compact4}) and (\ref{bkm:eq_compact5}). \\  } 
\centering
\begin{tabular}{ccr}
\toprule
	\textbf{Equations formulation} & ~~~ & \textbf{Cost estimate $-\xi_1/\lambda_1$} \\
\midrule
	(17), (18) & & 0.794  \\ 
	(17), (18.a) & & 0.767 \\ 
	(17.a), (18) & & 1.027 \\ 
	(17.a), (18.a) & & 1.000  \\ 
\bottomrule
\end{tabular}
\label{bkm:tbl1}
\end{table}

\section{Discussion and conclusion}

The example optimization problem analysed in this paper is rather simple as compared to e.g. DICE model~\cite{refNordhaus2019}, \cite{refDICEWebPage}. 
However, even that simple problem appears to be analytically intractable.
A numerical solution of that problem can be easily found with sufficiently high precision and FOC can be checked numerically to ensure the validity of the obtained solution.
The marginal cost of the optimal control can be analytically expressed and calculated.
The "cost estimate" $x$ obtained from the concept of matching marginal quantities added to problems' equations and preserving the optimal utility, that is, an optimal cost estimation method employing a ratio of two marginal values (Lagrange multipliers) leads to an estimate clearly deviating from the true cost of optimal control. Moreover, such an estimate is ill-defined, because it's arbitrary (non-unique) and depends on a particular way of expressing exactly the same problem i.e. a formulation that for all controls (inputs) delivers the same trajectories (outputs). 

While the considered example is rather hypothetical i.e. the overall model and particular functions employed to describe damage to fruit, flies reproduction, and economic cost of pesticides application were not validated, and the measurement units (e.g. tons of flies) might sound awkward, the purpose of this theoretical ad-hoc construction is to demonstrate a setting in which shadow prices are being used in practice. 

The semi-analytical example provided here substantially limits the application of marginal values (shadow prices) for optimal cost estimate approaches similar to (\ref{bkm:eq_margInc}) including the SCC calculation approach implemented in the DICE model~\cite{refSCCFrontiers}.


\section*{Acknowledgments} 

The authors acknowledge early discussions around the DICE model with their IIASA colleagues Elena Rovenskaya, Artem Baklanov, Fabian Wagner, Thomas Gasser, and Petr Havlik.
These discussions have spurred the interest to the application of shadow prices in DICE that ultimately resulted in the presented analysis.
The authors are grateful to their IIASA colleagues Johannes Bednar, Michael Kuhn, Stefan Wrzaczek, and Michael Freiberger for useful discussions and feedback.
The authors acknowledge William Nordhaus for making the DICE source code and documentation openly available as well as the clean and transparent model structure that all made the present research possible. 
The authors acknowledge arXiv for hosting the earlier work on this topic~\cite{refSCCPrePrint}.

\section*{Funding} 

Austrian Science Fund (FWF): P31796-N29/``Medium Complexity Earth System Risk Management'' (ERM).

\section*{Author contributions} 

NK has conceptualized the problem and carried out the investigation; MO has contributed to funding acquisition; NK, AS, and MO have discussed the results in the process of investigation; NK has drafted the paper; all co-authors have contributed to writing the manuscript.

\section*{Competing interests} 

Authors declare no competing interests. 






\begin{thebibliography}{99}

	\bibitem{refSmith1987} Alasdair Smith. “Shadow Price Calculations in Distorted Economies.” The Scandinavian Journal of Economics, vol. 89, no. 3, 1987, pp. 287–302. JSTOR, https://doi.org/10.2307/3440199. Accessed 2 Sep. 2022.

	\bibitem{refStarrett2000} David A. Starrett. “Shadow Pricing in Economics.” Ecosystems, vol. 3, no. 1, 2000, pp. 16–20. JSTOR, http://www.jstor.org/stable/3658662. Accessed 2 Sep. 2022.


	\bibitem{refDanWu2021} Dan Wu, Shuwei Li, Li Liu, Jiyao Lin, Shiqiu Zhang,
Dynamics of pollutants’ shadow price and its driving forces: An analysis on China’s two major pollutants at provincial level,
Journal of Cleaner Production,
Volume 283,
2021,
124625,
ISSN 0959-6526,
https://doi.org/10.1016/j.jclepro.2020.124625.

	\bibitem{refElvira2022} Elvira Silva, Manuela Magalhães,
Environmental Efficiency, Irreversibility and the Shadow Price of Emissions,
European Journal of Operational Research,
2022,
ISSN 0377-2217,
https://doi.org/10.1016/j.ejor.2022.08.011.

	\bibitem{refZhaohua2022} Zhaohua Wang, Yanwu Song, Zhiyang Shen,
Global sustainability of carbon shadow pricing: The distance between observed and optimal abatement costs,
Energy Economics,
Volume 110,
2022,
106038,
ISSN 0140-9883,
https://doi.org/10.1016/j.eneco.2022.106038.

	\bibitem{refJunfu2022} Junfu Zhang,
JUE Insight: Measuring the Stringency of Land Use Regulation Using a Shadow Price Approach,
Journal of Urban Economics,
2022,
103461,
ISSN 0094-1190,
https://doi.org/10.1016/j.jue.2022.103461.

	\bibitem{refAslihan2011} Aslıhan Arslan,
Shadow vs. market prices in explaining land allocation: Subsistence maize cultivation in rural Mexico,
Food Policy,
Volume 36, Issue 5,
2011,
Pages 606-614,
ISSN 0306-9192,
https://doi.org/10.1016/j.foodpol.2011.05.004.

	\bibitem{refXiaojie2022} Xiaojie Wen, Shunbo Yao, Johannes Sauer,
Shadow prices and abatement cost of soil erosion in Shaanxi Province, China: Convex expectile regression approach,
Ecological Economics,
Volume 201,
2022,
107569,
ISSN 0921-8009,
https://doi.org/10.1016/j.ecolecon.2022.107569.

	\bibitem{refMichiyuki2018} Michiyuki Yagi, Shunsuke Managi,
Shadow price of patent stock as knowledge stock: Time and country heterogeneity,
Economic Analysis and Policy,
Volume 60,
2018,
Pages 43-61,
ISSN 0313-5926,
https://doi.org/10.1016/j.eap.2018.09.001.

	\bibitem{refAlexandros2020} Alexandros Maziotis, Andres Villegas, María Molinos-Senante,
The cost of reducing unplanned water supply interruptions: A parametric shadow price approach,
Science of The Total Environment,
Volume 719,
2020,
137487,
ISSN 0048-9697,
https://doi.org/10.1016/j.scitotenv.2020.137487.

	\bibitem{refKai2022} Kai Hou, Puting Tang, Zeyu Liu, Hongjie Jia, Kai Yuan, Chongbo Sun, Yi Song,
A fast optimal load shedding method for power system reliability assessment based on shadow price theory,
Energy Reports,
Volume 8, Supplement 1,
2022,
Pages 352-360,
ISSN 2352-4847,
https://doi.org/10.1016/j.egyr.2021.11.104.

	\bibitem{refFrancesca2019} Francesca Mariani, Maria Cristina Recchioni, Mariateresa Ciommi,
Merton’s portfolio problem including market frictions: A closed-form formula supporting the shadow price approach,
European Journal of Operational Research,
Volume 275, Issue 3,
2019,
Pages 1178-1189,
ISSN 0377-2217,
https://doi.org/10.1016/j.ejor.2018.12.022.

\bibitem{refNordhaus2019} William Nordhaus, Climate Change: The Ultimate Challenge for Economics, American Economic Review 2019, 109(6), 1991--2014, DOI:10.1257/aer.109.6.1991

\bibitem{refDICEWebPage} Documentation and source code for the DICE model are available online at https://williamnordhaus.com/dicerice-models where the direct link to the DICE User's Manual is: \\ http://www.econ.yale.edu/\~nordhaus/homepage/homepage/ \\ documents/DICE\_Manual\_100413r1.pdf


\bibitem{refSimonBlume1994} C. Simon and L. Blume, Mathematics for Economists, W. W. Norton \& Company, New York, 1994.

\bibitem{refFeichtinger2008} Dieter Grass, ed. Optimal Control of Nonlinear Processes: With Applications in Drugs, Corruption, and Terror. Berlin: Springer, 2008. ISBN: 978-3-540-77646-8.

\bibitem{refSCCPrePrint} Nikolay Khabarov, Alexey Smirnov, Michael Obersteiner, Social Cost of Carbon: What Do the Numbers Really Mean? ArXiv; DOI: 10.48550/arXiv.2001.08935

\bibitem{refSCCFrontiers} Nikolay Khabarov, Alexey Smirnov and Michael Obersteiner (2022), Social cost of carbon: A revisit from a systems analysis perspective. Front. Environ. Sci. 10:923631. DOI:10.3389/fenvs.2022.923631

\end{thebibliography}
\end{document}